\documentstyle[psfig]{aipproc}

\begin{document}
\title{Gamma-ray and X-ray luminosities from spin-powered pulsars\\
 in the full polar cap cascade model}

\author{Bing Zhang and Alice K. Harding}
\address{Laboratory of High Energy Astrophysics, NASA Goddard Space
 Flight Center.}

\maketitle

\begin{abstract}
We modify the conventional curvature radiation (inverse Compton 
scattering) + synchrotron radiation polar cap cascade model by
including the inverse Compton scattering of the higher generation 
pairs. Within the framework of the space-charge-limited-flow 
acceleration model with frame-dragging proposed by Harding \& 
Muslimov (1998), such a full polar cap cascade scenario can well 
reproduce the $L_\gamma \propto (L_{\rm sd})^{1/2}$ and the 
$L_x \sim 10^{-3} L_{\rm sd}$ dependences observed from the known
spin-powered pulsars. According to this model, the ``pulsed'' soft 
ROSAT-band X-rays from most of the millisecond pulsars might be of 
thermal origin, if there are no strong multipole magnetic
components near their surfaces.
\end{abstract}

\section*{Introduction}

Eight and 35 spin-powered pulsars have been also detected in
$\gamma$-ray and X-ray bands, respectively. Despite their
great diversity of emission features, the luminosities of
these pulsars seem to obey the empirical laws $L_\gamma\propto 
(L_{\rm sd})^{1/2}\propto B/P^2$ (Thompson et al. 1997), and $L_{x}
\sim 10^{-3} L_{\rm sd}\propto B^2/P^4$ (Becker \& Tr\"umper 1997), 
where $L_{\rm sd}$ is the spin-down luminosity of the pulsar. The
spectra of the $\gamma$-ray emission and the X-ray emission
from most of the pulsars are non-thermal, while full surface thermal 
emission components are identified from Vela, Geminga, PSR 1055-52 and 
PSR 0656+14, and possible hot polar cap thermal emission components are
reported from PSR 1929+10 and PSR J0437-4715.

Two competing models for pulsar high energy emission, i.e., the polar cap 
models (Daugherty \& Harding 1996; Sturner, Dermer \& Michel 1995) and 
the outer gap models (Cheng, Ho \& Ruderman 1986; Zhang \& 
Cheng 1997) were proposed. Canonical polar cap cascade models involve 
the curvature radiation (CR) or inverse Compton scattering (ICS) of the 
primary particles and the synchrotron radiation (SR) of higher generation 
pairs. Here we modify such a cascade picture by including the ICS of the 
higher generation pairs, which is important since it usually occurs in
the resonant regime. A more detailed presentation of this study is shown
in Zhang \& Harding (1999).

\section*{The Model} 
 
\subsection*{The ``full-cascade'' picture}

The ``full-cascade'' scenario is: primary particles accelerated from
the inner gap emit primary $\gamma$-rays via CR or ICS, these 
$\gamma$-rays will pair produce in strong
magnetic fields. The secondary pairs have non-zero pitch angle with
respect to the field lines. The perpendicular energy of the pairs will
go to high energy radiation via SR, and the parallel energy of the 
pairs will also convert to radiation via ICS with the soft thermal
photons from either the full neutron star surface or the hot polar cap. 
Under the condition of $\gamma\ge\gamma_{\rm res}={48 B_{12}{\rm Max}
[(1-\beta\mu_{\rm s,max})T_{s,6}, (1-\beta\mu_{\rm h,max})T_{h,6}]^{-1}}$
(i.e. the ``resonant scattering condition'', Dermer 1990), the 
efficiency of converting particles' kinetic energy to radiation by 
ICS is almost 100\%, so that the total high energy emission luminosity
is approximately the polar cap particle luminosity. Here $T$ is
the temperature of the soft photons, and $\mu_{\rm max}$ is the cosine
of the maximum scattering angle. Note that two components, i.e., a
soft full surface thermal emission (denoted by `s') and a hard hot 
polar cap thermal emission (denoted by `h'), are adopted.
 
The basic ingredients in constructing an analytic description of such
a full cascade scenario are (1) the energy distribution in SR 
(perpendicular) and ICS (parallel) branches, and (2) the recursion 
relations between different generations. For the former, 
$\eta_\parallel= {\gamma_{i,\parallel} / \gamma_i} =
{[1+(\gamma_{i+1}^2-1)\sin^2 \theta_{\rm kB}]^{-1/2}}$ 
and $\eta_\perp=1-\eta_\parallel$ are the energy portions for the 
parallel (ICS) and the perpendicular (SR) branches, respectively,
where $\gamma_i$ and $\gamma_{i,\parallel}$ are the total and parallel
Lorentz factors of the $i$-th generation pairs, and $\theta_{\rm kB}$
is the impact angle between the photon and the magnetic field line. 
With this, one can get the reduction factor of the typical photon 
energies for the adjacent generations, e.g., $\kappa_{\rm SR}=
{\epsilon_{i+1,{\rm SR}} / \epsilon_i}= (3/4)\chi$, 
and $\kappa_{\rm ICS}={\epsilon_{i+1,{\rm ICS}} / \epsilon_i}=
\eta_\parallel B'_e$, where $B'_e=B_e/B_{cri}$, and $\chi$ is the
parameter to describe the $\gamma-B$ pair production process. 
The photon escaping energy is $E_{\rm \gamma,esc}{\rm (nPSR)}\simeq 2.0
{\rm GeV}B_{e,12}^{-1}P^{1/2}r_{e,6}^{-1/2}\chi_{_{1/16}}$ 
for normal pulsars, and $E_{\rm \gamma,esc}{\rm (msPSR)}\simeq 
73{\rm GeV}B_{e,9}^{-1}P_{-3}^{1/2}r_{e,6}^{-1/2}\chi_{_{1/12}}$ 
for millisecond pulsars, where $\chi_{_{1/16}}=\chi/(1/16)$, and
$\chi_{_{1/12}}=\chi/(1/12)$, and $r_{e,6}$ is the emission height
in units of $10^6$cm. Given the typical primary photon energy 
$E_0$ (which is model-dependent), we can then get some non-integer
generation order parameters (Lu et al. 1994; Wei et al. 1997), 
e.g. $\zeta_{\rm SR}={\log(E_{\rm esc}/E_0) 
\over \log (\kappa_{\rm SR})}+1$ (for pure SR generations), and
$\zeta_{\rm ICS}={\log(E_{\rm esc}/E_0) \over \log (\kappa_{\rm ICS})}+1$ 
(for pure ICS generations), which can describe the complex cascade 
process analytically.

\subsection*{Harding \& Muslimov acceleration model} 

Harding \& Muslimov (1998) has improved the space-charge-limited flow 
acceleration model (Arons \& Scharlemann 1979) by incorporating the 
frame-dragging $E_\parallel$,
upper and lower pair formation front and both the CR and ICS 
of the primary electrons.
It was found that a stable accelerator is located at an effective 
``radius'' of $R_{\rm E}\sim (1.5-2) R$ for normal pulsars when ICS
energy loss is less than that of CR, since the ICS of the upward 
versus downward particles with the soft thermal photons are anisotropic due
to different geometries. The typical length of the CR-controlled gap
is $S_c \simeq 4.8\times 10^4 {\rm cm}B_{p,12}^{-4/7}P^{4/7} R_{E,6}
^{16/7} (\cos\alpha)^{-3/7}$, and the typical Lorentz factor of the 
primary particles is $\gamma_0=4.7\times 10^7 B_{p,12}^{-1/7} P^{1/7} 
r_{e,6}^{4/7} (\cos\alpha)^{1/7}$, where $\alpha$ denotes the 
inclination angle of the neutron star. Thus the typical energy of the 
primary photons is $E_0 \simeq 33.2 ({\rm GeV}) B_{p,12}^{-3/7}P^{-1/14}
r_{e,6}^{17/14}(\cos \alpha)^{3/7}$, with which one can get explicit 
expressions for the generation parameters, and complete the analytic 
description of the full-cascade model. 

\section*{Luminosity predictions} 
\subsection*{Gamma-ray luminosity} 
An interesting feature of the Harding \& Muslimov model is that 
$\gamma_0$ is insensitive to pulsar parameters, so that
the polar cap luminosity $L_{\rm pc}=\gamma_0 mc^2 \dot N_{ p}$ is
roughly proportional to $(L_{\rm sd})^{1/2}$. One advantage of 
the full cascade model is that the ICS branches can convert the
``lost'' parallel kinetic energies of the particles also to radiation,
so that the total high energy luminosity (mainly $\gamma$-ray luminosity)
is also roughly proportional to $(L_{\rm sd})^{1/2}$. More specifically,
the model predicts
\begin{mathletters}
\begin{eqnarray}
L_\gamma({\rm full})\simeq L_{\rm pc}&&=5.4\times 10^{31} {\rm erg\cdot
s^{-1}}B_{p,12}^{6/7} P^{-13/7}r_{e,6}^{4/7}(\cos\alpha)^{8/7}\\
&&=1.7\times 10^{16} B_{p,12}^{-1/7}P^{1/7}r_{e,6}^{4/7}
(\cos\alpha)^{8/7}(L_{\rm sd})^{1/2}.
\label{Lgamma2}
\end{eqnarray} 
\end{mathletters}
This nearly reproduces the observed $L_\gamma\propto (L_{\rm sd})^{1/2}$
feature. 

\subsection*{Thermal X-ray luminosity \label{th-x}} 
 
There are two thermal components in a pulsar's X-ray spectrum (though
they might be buried under the non-thermal component). For the full
surface thermal component, we have adopted a simple rough ``standard'' 
cooling model, which is not inconsistent with the observations.

For the hot polar cap thermal component, we treated it with the
self-consistent polar cap heating in the Harding \& Muslimov model.
Since the flow is space-charge-limited so that the deviation of
local charge density ($\rho$) from the Goldreich-Julian density 
($\rho_{\rm GJ}$) is small, the backflow particle luminosity should
be only a small portion (a factor of $\mid\rho-\rho_{\rm GJ}\mid/
2\rho_{\rm GJ}$) of the polar cap luminosity, which reads
\begin{equation} 
L_{x,th,{\rm max}}\simeq 5.9\times 10^{29}{\rm erg\cdot
s^{-1}}B_{p,12}^{2/7}P^{-9/7}r_{e,6}^{-22/7} (\cos\alpha)^{5/7},
\label{Lx-th} 
\end{equation} 
and the maximum polar cap temperature (assuming an area of $\pi r_p^2$, 
where $r_p=\theta_{\rm pc}R=1.45\times 10^4P^{-1/2}{\rm cm}$) is 
\begin{equation}
T_{\rm pc,max}=2.0 \times 10^6 {\rm K}
B_{p,12}^{1/14}P^{-1/14}r_{e,6}^{-11/14}(\cos\alpha)^{5/28}.
\end{equation}
 
\begin{figure}
\centerline{\psfig{file=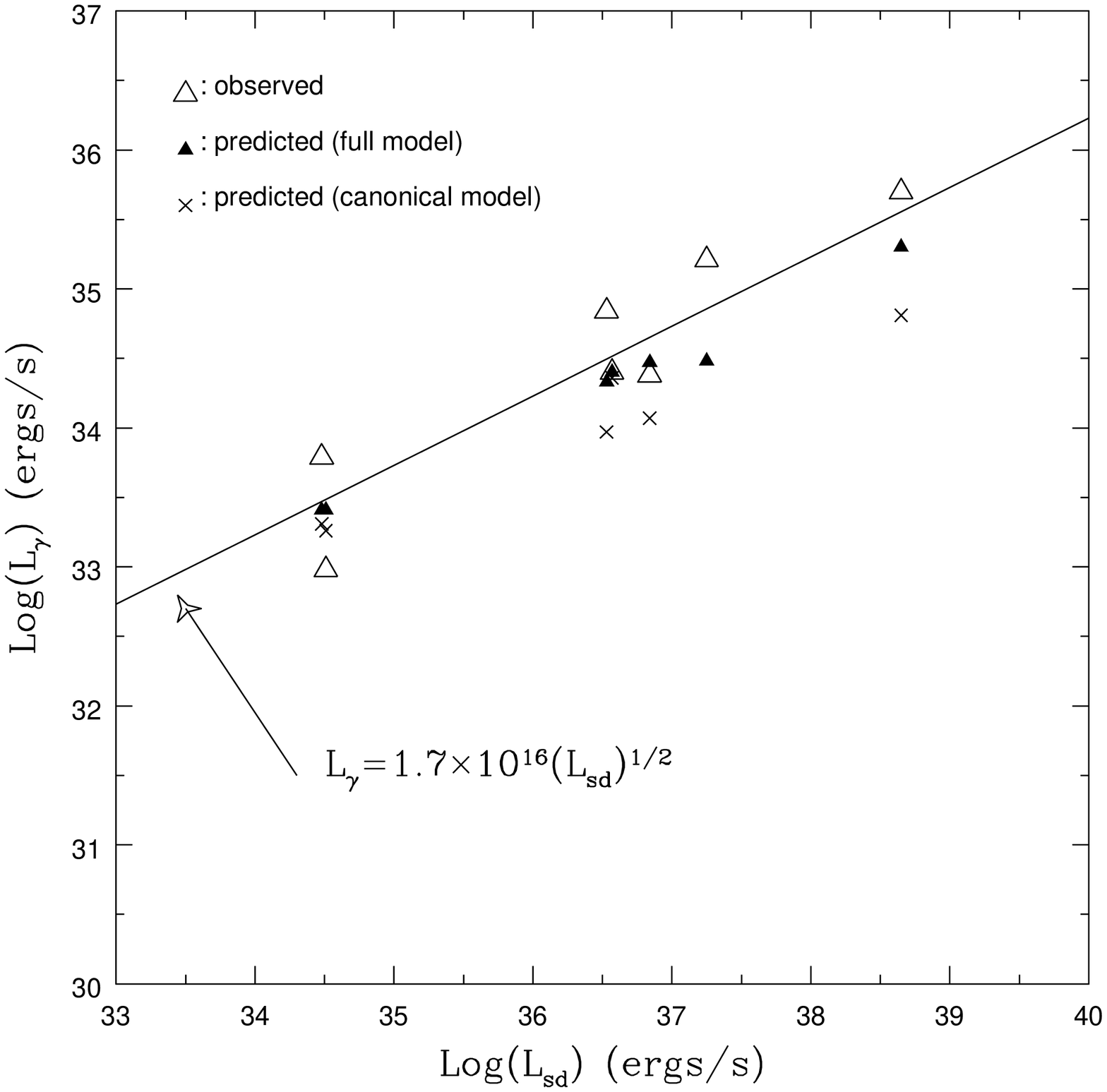,height=3.0in,width=3.0in}
 \psfig{file=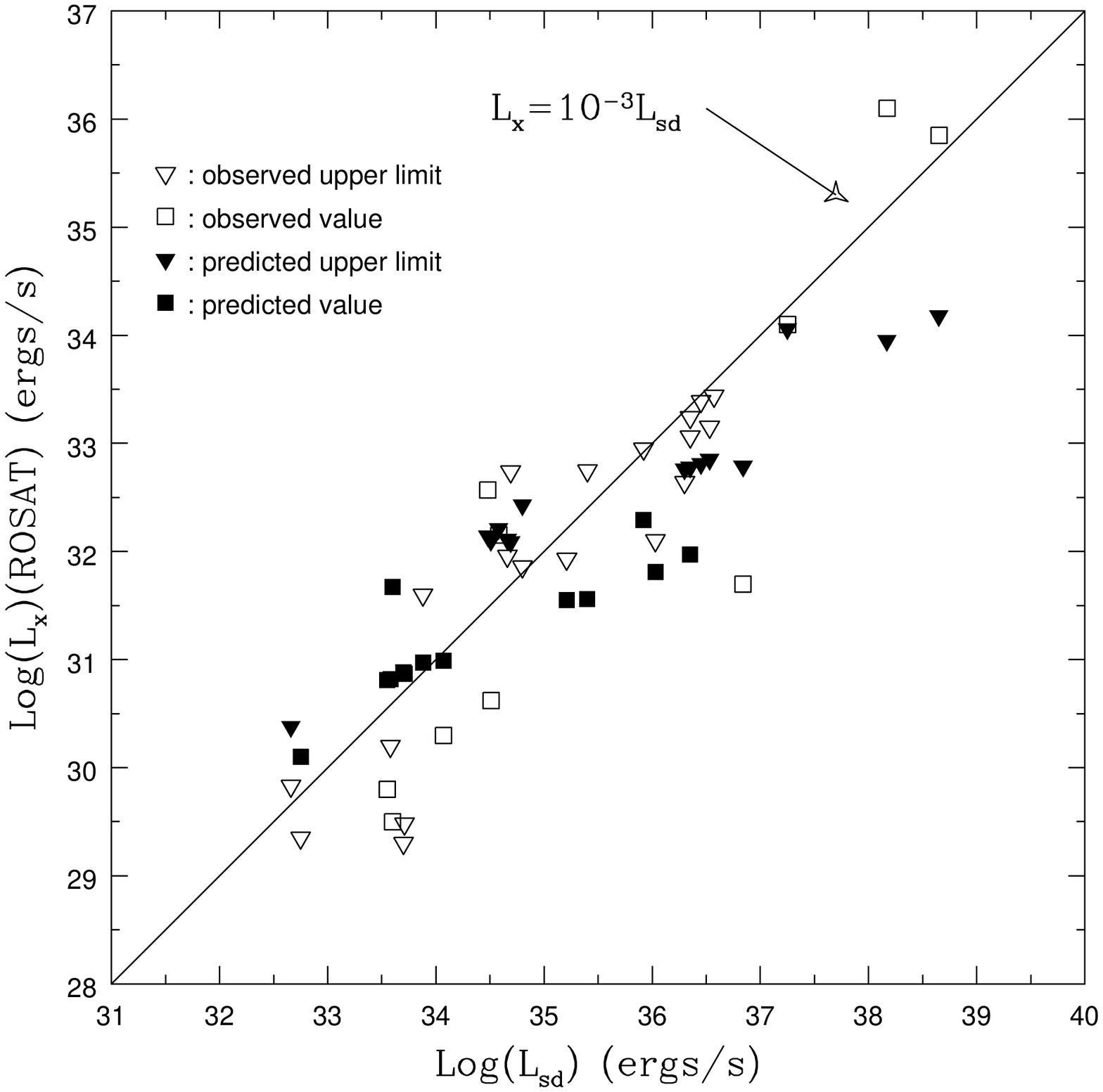,height=3.0in,width=3.0in}}
\vspace{10pt}
\caption{Observation versus theory: (a) $\gamma$-ray luminosities
 (b) X-ray luminosities.}
\end{figure}


\subsection*{Non-thermal X-ray luminosity \label{nonth-x}} 
The SR spectra of all SR branches can not get down to X-ray energies 
as those observed by ROSAT and ASCA, since there is a low
energy cut-off at the blueshifted local resonant frequency
(corresponding to the transition between the ground and the first
excited Landau levels). Another advantage of
the full cascade scenario is that, the ICS branches, which were
neglected in the canonical cascade model, can naturally give a
non-thermal component extend to the X-ray band. For normal pulsars,
the maximum non-thermal luminosity below a certain energy $E_c$ 
could be estimated as
\begin{equation} 
L_{x,nth}(E_c)\le L_{\rm pc}\sum_{k=1}^{\rm int(\zeta_{\rm
SR})} \left[\eta_\perp^{k-1} \left(\sum_{j=1}^{\rm 
int(\zeta_{\rm ICS,k-1})}\eta_\parallel^{j} \eta_{c,k,j} 
\right)\right],
\label{Lxi} 
\end{equation} 
where $\zeta_{\rm ICS,k}={\log(E_{\rm esc}/E_k) \over
\log(\kappa_{\rm ICS})}+1$ is the number of pure ICS generations 
for the typical energy of the $k$-th SR generation, $E_k=E_0 
\kappa_{\rm SR}^{k}$, to reduce to the escaping energy $E_{\rm esc}$,
$\eta_{c,k,j}= \gamma_c/\gamma_{k,j}$ (but $=1$ when $\gamma_c 
\geq \gamma_{k,j}$), $\gamma_c=E_c/[(1-\beta\mu)2.8 kT]$, and
$\gamma_{k,j}=(\epsilon_0/2)\kappa_{\rm SR}^{k-1} \kappa_{\rm ICS}^{j-1}
 \eta_\parallel$. For millisecond pulsars, a slightly
different formula is adopted (see Zhang \& Harding 1999). When 
calculating the X-ray luminosity within a certain band, $E_{c1}
<E<E_{c2}$, we then have $L_{\Delta E}=L_{x,nth}(E_{c2}) -L_{x,nth}
(E_{c1})$.


\section*{Results and discussions} 
The observation versus model prediction of the $\gamma$-ray and 
X-ray pulsars are shown in Fig.1 and Fig.2. For the X-ray luminosities,
three components (the non-thermal, the full surface thermal and the
hot polar cap thermal components) are taken into account. 

An obvious conclusion is that the full polar cap cascade model 
within the framework of Harding \& Muslimov acceleration model can 
both reproduce the $L_{\gamma}\propto (L_{\rm sd})^{1/2}$ and
$L_x(ROSAT)\sim 10^{-3} L_{\rm sd}$ feature simultaneusly. The former 
was not done by the outer gap model which interprets non-thermal X-ray
emission as the SR of the downward cascade from the outer gap
(Cheng \& Zhang 1999; Zhang \& Cheng 1997). In our model, we also
compare the non-thermal X-ray luminosity with the luminosities
of the two thermal components. It is found that for middle-aged 
pulsars such as Vela, Geminga, PSR 1055-52, and PSR 0656+14, the 
full surface thermal luminosity is comparable to the non-thermal 
one, so that such a component should be detectable from these 
pulsars. Such a feature is actually 
observed. For the hot polar cap thermal component, our model shows
that it is detectable in relative old pulsars such as PSR 0950+08
and PSR 1929+10, although the non-thermal component is also
detectable. The outer gap model actually only predicts pure thermal
components in these pulsars, since the thick outer gap does not exist
in their magnetospheres (Cheng \& Zhang 1999). For the millisecond 
pulsars, our model predicts that the thermal emission from polar cap
heating is the dominant component in the ``pulsed'' ROSAT-band
spectra, while the outer gap model predicts non-thermal emission, though
they assume strong multipole magnetic fields near the surfaces of the
millisecond pulsars. Future observations and spectral analyses can
distinguish the riveling models.

\end{document}